\documentclass{PoS}

\def\be{\begin{equation}}
\def\ee{\end{equation}}
\usepackage{amsmath}

\title{On a relation between production processes and total cross sections}

\ShortTitle{On a relation between production processes and total cross sections}

\author{\speaker{St\'ephane MUNIER}\\
        (Centre de physique th\'eorique, CNRS, \'Ecole polytechnique, Palaiseau, France)\\
        E-mail: \email{stephane.munier@cpht.polytechnique.fr}}

\abstract{
Perturbative QCD is the appropriate tool to describe many important properties of the inclusive observables
measured at electron-proton (or ion) colliders, such as the energy dependence of the total cross sections in
well-chosen kinematical regions. This is because the electron may effectively be replaced by its cloud of
photons, whose virtualities provide a hard scale that enables perturbative expansions.

At hadron colliders instead, there is no hard scale in the initial state. Therefore, 
the observables one may
compute perturbatively involve the production of jets, and thus belong to a quite different 
class of observables.

However, it turns out that there is a formal relation between production processes and total cross sections,
enabling one to apply calculations of the latter to the former. We review this relation, 
and present our recent proof that it holds at next-to-leading order (in the BFKL sense).
}

\FullConference{XXI International Workshop on Deep-Inelastic Scattering and Related Subjects\\
22-26 April, 2013\\
Marseille, France.\footnote{Also presented
at the ECT$^*$ workshop on high energy, high density and hot QCD,
17-21 June, 2013, Trento, Italy.}}

\begin{document}

\section{Introduction}

Parton densities "seen" in hadronic collisions increase with the energy of
the collisions. This growth is predicted by linear evolution equations such
as the BFKL equation, established in QCD.
At very high energies, parton densities may become so large that they
saturate, which means that the evolution equations become nonlinear
(and change name: they become the BK and B-JIMWLK equations, see 
Ref.~\cite{Gelis:2010nm}
for a review). These new equations predict in particular
the emergence of a hard, energy-dependent, momentum scale
called the saturation scale.
This regime is very interesting theoretically. Parton saturation may also
have important phenomenological consequences at the LHC.

The question is how to test this exciting regime of QCD.
Electron-hadron collisions, 
happening through the exchange of a virtual photon,
may be understood as a dipole of "tunable" size $r$, of the order of the
inverse virtuality of the exchanged photon, scattering off the
hadron (in practice, proton or nucleus).
A lot of understanding of the dipole scattering amplitude was gained at HERA,
at the border of the dense regime of QCD.
On the theoretical side, it is "easy" to formulate the QCD evolution
of the dipole amplitude with the energy as radiative corrections to
the dipole wave function. One arrives at the BFKL equation in the regime
of low densities, and at the BK and B-JIMWLK equations if one tries
to account for high-density effects.

\begin{figure}[h]
\begin{center}
\includegraphics[width=6cm]{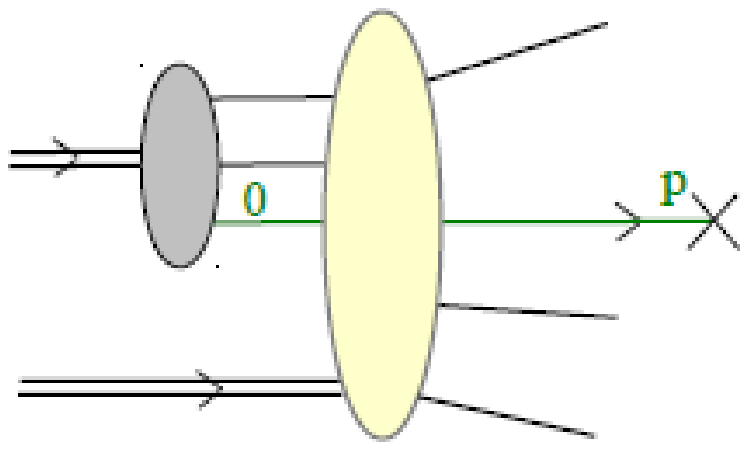}
\hskip 2cm
\includegraphics[width=6cm]{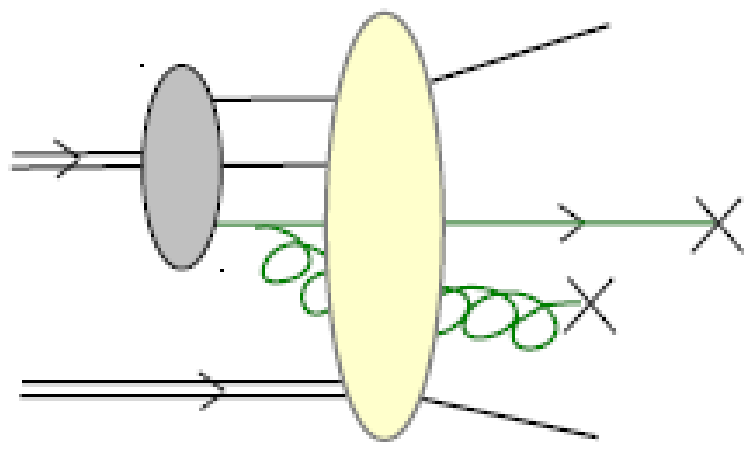}
\end{center}
\caption{\label{fig0}Schematic view of
$p_\perp$-broadening (left; a jet
of transverse momentum $p_\perp$ is measured) and dijet correlations
(right; two forward jets are measured, separated
by some azimuthal angle $\Delta\phi$).}
\end{figure}

At a hadron collider instead, where there is no hard scale such as the
photon virtuality
in the initial state, one needs to find appropriate production processes.
The simplest of those may be the so-called "$p_\perp$-broadening" 
process (see Fig.~\ref{fig0}, left) in which 
one observes in the final state a jet of transverse momentum
$p_\perp$ together with an arbitrary number of other particles. 
The interpretation is the following: Through the interaction
with hadron number 2, one valence quark
of hadron number 1 acquires a momentum of the order of the saturation scale,
which may be large if hadron 2 is a large nucleus and/or 
if the energy of the interaction is high.

There are other more 
sophisticated
observables that have been discussed in the literature.
The energy dependence of forward 
dijet azimuthal correlations (see Fig.~\ref{fig0}, right)
is one of them \cite{Marquet:2007vb}.

An outstanding challenge is to take over what has been learnt at HERA
to the LHC to push further our understanding of dense QCD.
A priori, this looks very hard.
However, there exists a surprising relation between total cross
sections like the ones measured in deep-inelastic scattering, and
production processes measured at hadron colliders.
This relation was established at the classical level some time ago.
Many authors have then assumed that it is true also when quantum corrections
are taken into account, but no proof was available beyond leading order
until recently. It is the purpose of this contribution to report
on the recent progress in putting this relation on solid ground~\cite{Mueller:2012bn}.

We have studied in detail $p_\perp$-broadening, but we believe that
all our analysis of quantum corrections may go over to 
other production processes.


\section{Formulation of a production process in proton-nucleus collision}

Let us write the rate of production
of a jet of transverse momentum $p_\perp$ in quark-nucleus scattering:
\be
\frac{dN}{d^2p_\perp}=\int \frac{d^2x_\perp}{(2\pi)^2}
e^{-i p_\perp x_\perp}
\times\left[\frac{1}{N_c}\text{tr}
\left\{
\sum_n\langle n|V_{0_\perp}|A\rangle^*\langle n|V_{x_\perp}|A\rangle
\right\}\right]
\label{eq:correspondence}
\ee
where $V$ is the usual Wilson line which represents the 
quark propagating in the classical
field of the nucleus, and $\left| A\right\rangle$
is a nuclear state.
In the McLerran-Venugopalan model\footnote{%
The question whether this equivalence would be true for a general
interaction, beyond the simplifying assumptions of the
McLerran-Venugopalan model
on the statistical properties of the field of the 
nucleus, was raised by several participants
to the workshop. The answer is not known. It seems crucial for our
calculation that the gluons exchanged with the target
be at most pairwise correlated, but we cannot exclude that
our calculation would eventually turn out to be more general.} 
\cite{McLerran:1993ni},
the factor in the square brackets can be straightforwardly
identified with the
$S$-matrix element for the elastic scattering of a color dipole of
size $x_\perp$ off a large
nucleus, denoted by $S(x_\perp)$.
\begin{figure}
\begin{center}
\includegraphics[width=8cm]{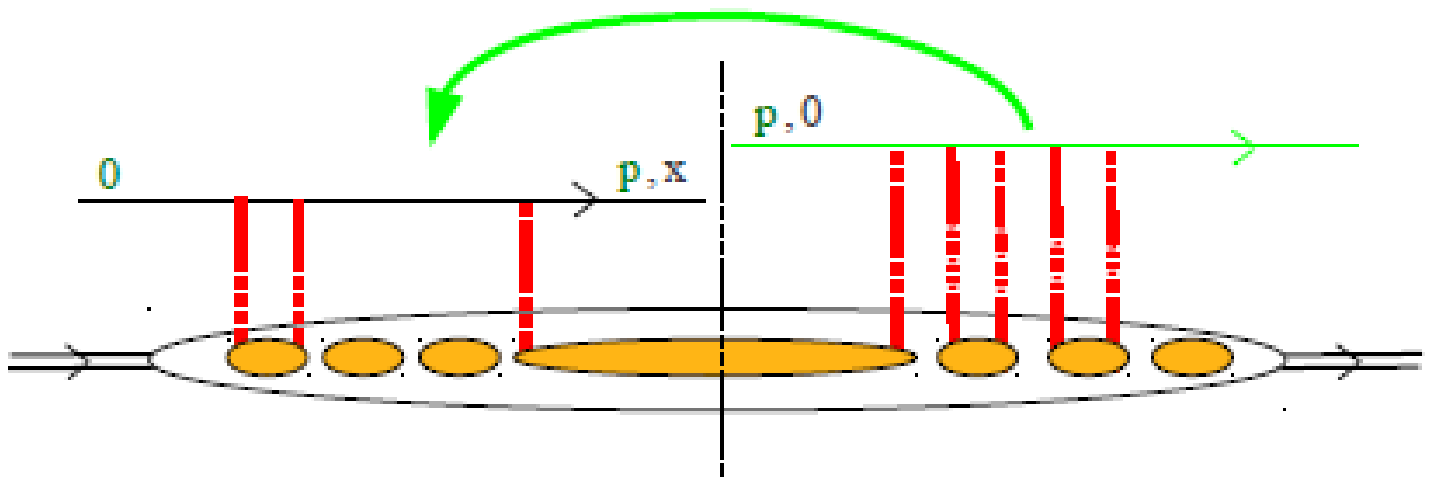}
\includegraphics[width=6.5cm]{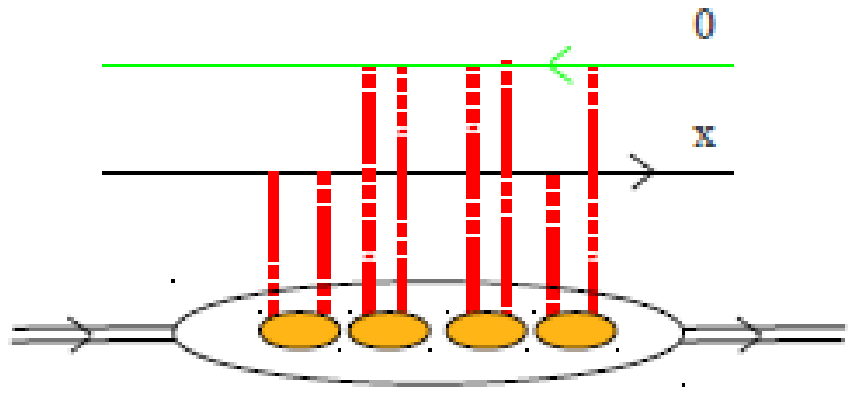}
\end{center}
\caption{\label{corr1}Graphical illustration of the correspondence
between the rate of production of a jet
of transverse momentum $p_\perp$ in $pA$ collisions and
the dipole scattering amplitude off a nucleus in the McLerran-Venugopalan
model. The red vertical lines are exchanged gluons.}
\end{figure}

So we arrive at a surprising relation between two seemingly very
different processes: {\em The rate of production of a jet
of transverse momentum $p_\perp$ is the 2-dimensional Fourier transform of the $S$-matrix
element for the elastic scattering of a color dipole whose size $x_\perp$ is conjugate to
$p_\perp$.}

This identification can be illustrated by an appealing drawing
(see Fig.~\ref{corr1}): The quark in the complex-conjugate amplitude
of the broadening process may be "bent over" to become an antiquark
in the amplitude, and thus, together with the quark in the amplitude,
form a color dipole. This picture was presumably first proposed 
by Zakharov \cite{Zakharov:1996fv}, in a different context however.
While this picture
is literally true in the McLerran-Venugopalan model, it is
not clear whether it would be true beyond the
classical approximation.

When quantum
corrections are included, the factor $S(x_\perp)$ represented by
the square brackets in
Eq.~(\ref{eq:correspondence}) should be
changed to
\be
\tilde S(x_\perp)=\frac{1}{N_c}
\sum_n
\text{tr}
\bigg\{
\langle n|T\left(V_{0_\perp}e^{i\int d^4y {\cal L}_I(y)}\right)
|A\rangle^*
\langle n|T\left(V_{x_\perp}e^{i\int d^4y {\cal L}_I(y)}\right)
|A\rangle
\bigg\},
\label{eq:Stilde}
\ee
where ${\cal L}_I$ is the QCD interaction Lagrangian,
say in the lightcone gauge.
But now, it is not clear that the identification with
a dipole $S$-matrix element may still hold. Indeed, the
latter would read, when including quantum corrections,
\be
S(x_\perp)=\frac{1}{N_c}\text{tr}
\langle A|
T\left(V_{0_\perp}^\dagger V_{x_\perp} e^{i\int d^4y {\cal L}_I(y)}\right)
|A\rangle,
\label{eq:Sdipolenucleus}
\ee
and a priori, 
there is no reason why this $S(x_\perp)$ should be equal to $\tilde S(x_\perp)$.

\begin{figure}
\begin{center}
\includegraphics[width=8cm]{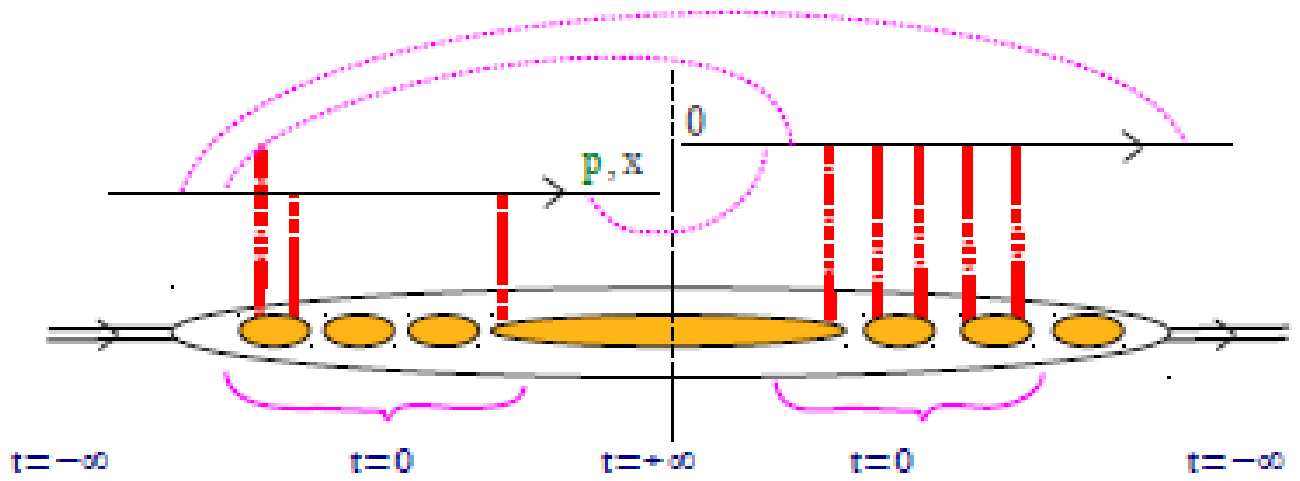}
\includegraphics[width=6.5cm]{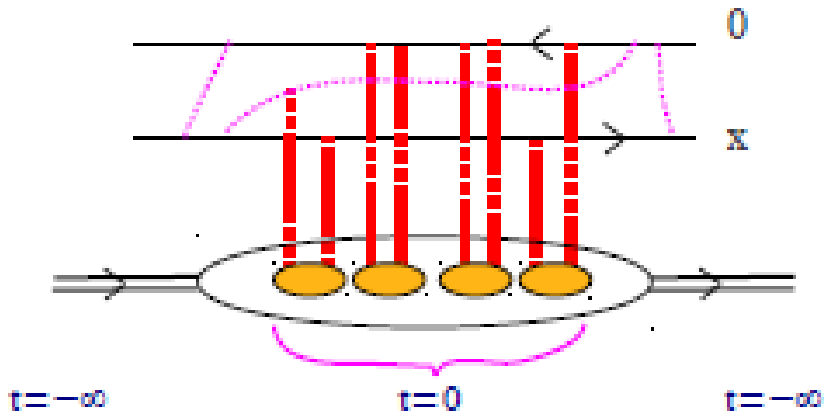}
\end{center}
\caption{\label{corr2}
Illustration of the would-be 
identity between jet-production rate and dipole
cross section beyond the McLerran-Venugopalan model.
The pink lines represent gluonic fluctuations of the quark.
}
\end{figure}

In the next sections, we are going to report on our analysis 
of these two processes in
perturbation theory, in view of trying and establishing an
identity between them beyond the McLerran-Venugopalan model.
Needless to say, we will not be able to provide any detail:
We shall refer the interested reader to the original 
paper~\cite{Mueller:2012bn}.


\section{Quantum corrections: leading order}

\begin{figure}
\begin{center}
\includegraphics[width=15cm]{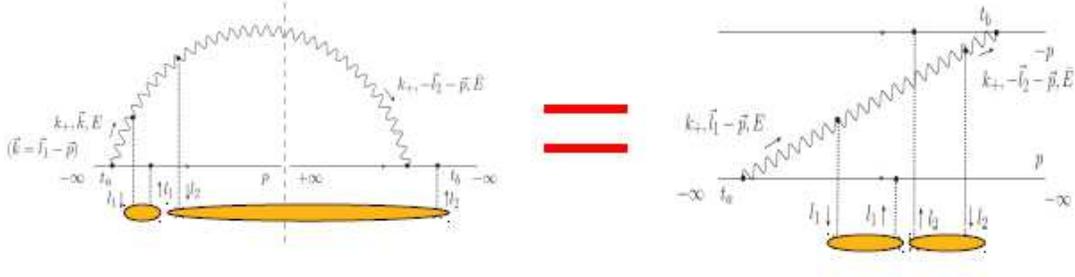}
\end{center}
\caption{\label{lo1}
Simplest graph contributing to the broadening cross section (left) 
and the corresponding graph contributing to the elastic dipole amplitude (right).
}
\end{figure}

In this section, we consider the first quantum correction to
the McLerran-Venugopalan model.
We pick one graph on the broadening side and the would-be
corresponding one on the
dipole side (see Fig.~\ref{lo1}). 
On the broadening side,
it is an interference graph
between the emission of a gluon in the initial state in the amplitude, and 
the emission of a gluon in the final state in the complex-conjugate amplitude.
Using the rules of time-ordered perturbation theory,\footnote{
The rules are listed for example on the recent textbook of Ref.~\cite{KL},
although we use slightly different conventions.} the evaluation
of this graph is especially simple since the quark-gluon vertices
are all treated in the eikonal approximation. We find that the
particular broadening graph drawn in Fig.~\ref{lo1} has the expression
\begin{multline}
\left.\frac{dN}{d^2p}\right|_{
\text{\scriptsize graph
in Fig.~\ref{lo1}}}
=-\frac{\alpha_s N_c}{N_c^2-1}\int_0^{+\infty}\frac{dk_+}{k_+}
\int \frac{d^2\vec l_{1}}{\vec l_{1}^2} \frac{d^2\vec l_{2}}{\vec l_{2}^2}
\frac{(\vec p-\vec l_{1})\cdot(\vec p+\vec l_{2})}
{(\vec p-\vec l_{1})^2(\vec p+\vec l_{2})^2}\\
\times\left[
\alpha_s x g(x,\vec l_{1}^2)\rho L
\right]
\left[
\alpha_s x g(x,\vec l_{2}^2)\rho L
\right],
\end{multline}
where we have used standard notations.
Here we have included two particular
exchanges of pairs of gluons with the target,
one elastic and one inelastic scattering,
represented by the factors in the last line of the previous equation.
But it would be straightforward to include an arbitrary number of exchanges,
provided that the $t$-channel gluons have pairwise correlations.
The evaluation of the dipole graph in Fig.~\ref{lo1} gives exactly
the same result, hence there is a perfect correspondence 
between these graphs, even before the integration over the momenta, provided
that one labels properly the latter.

\begin{figure}
\begin{center}
\includegraphics[width=4cm]{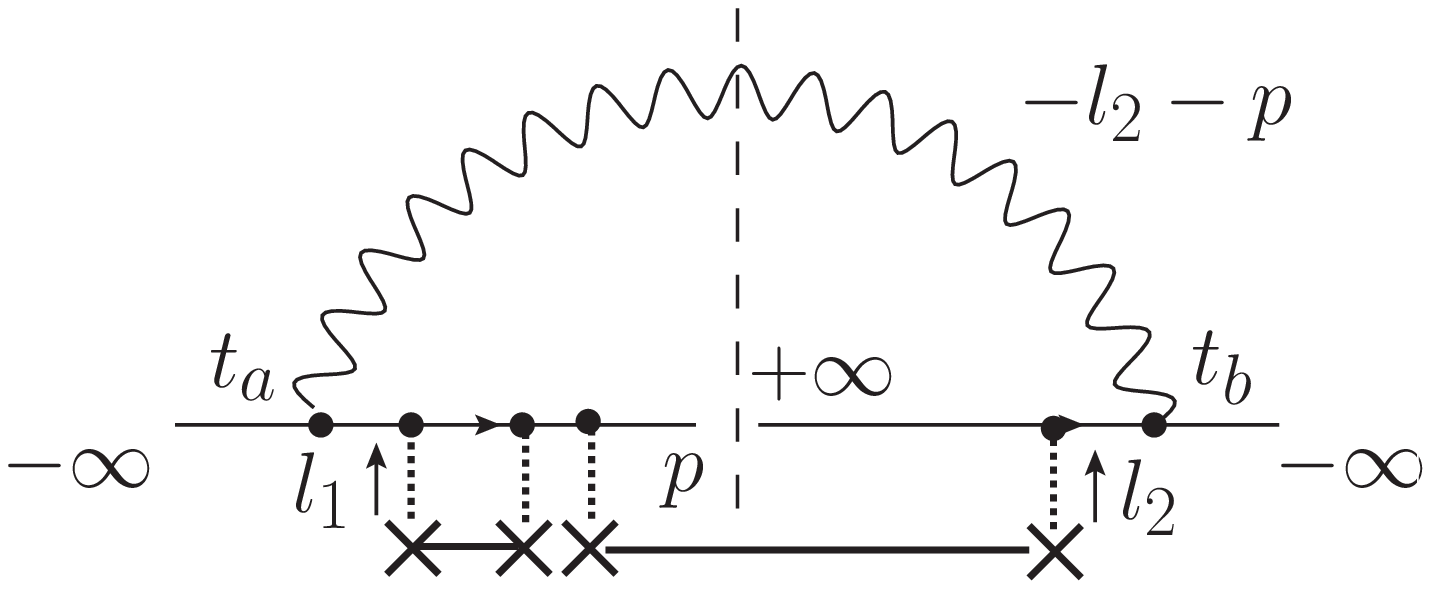}
\includegraphics[width=1.5cm]{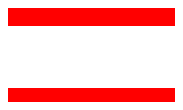}
\begin{tabular}{cc}
\includegraphics[width=4cm]{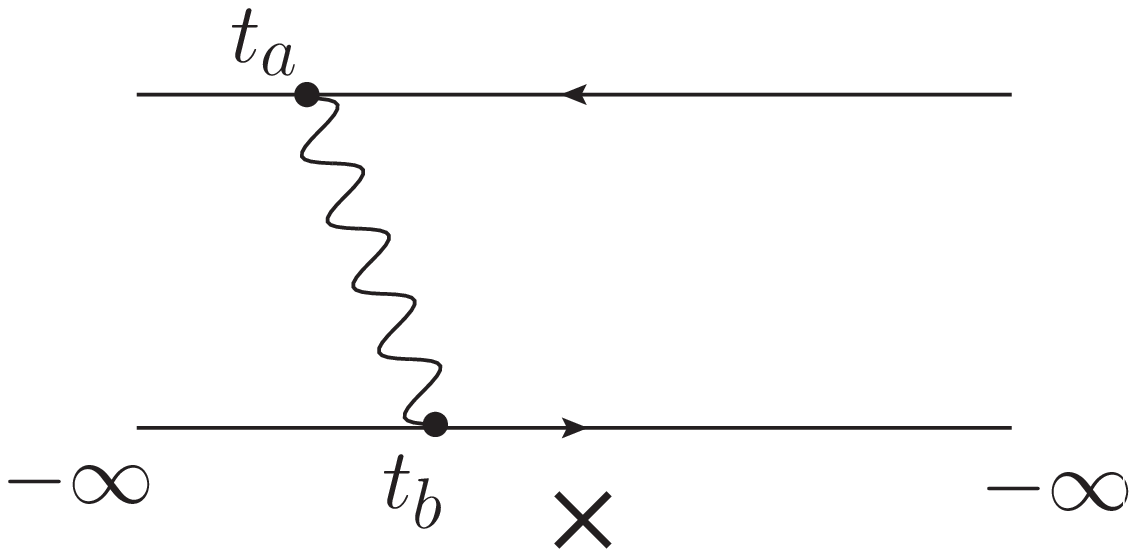} &
\includegraphics[width=4cm]{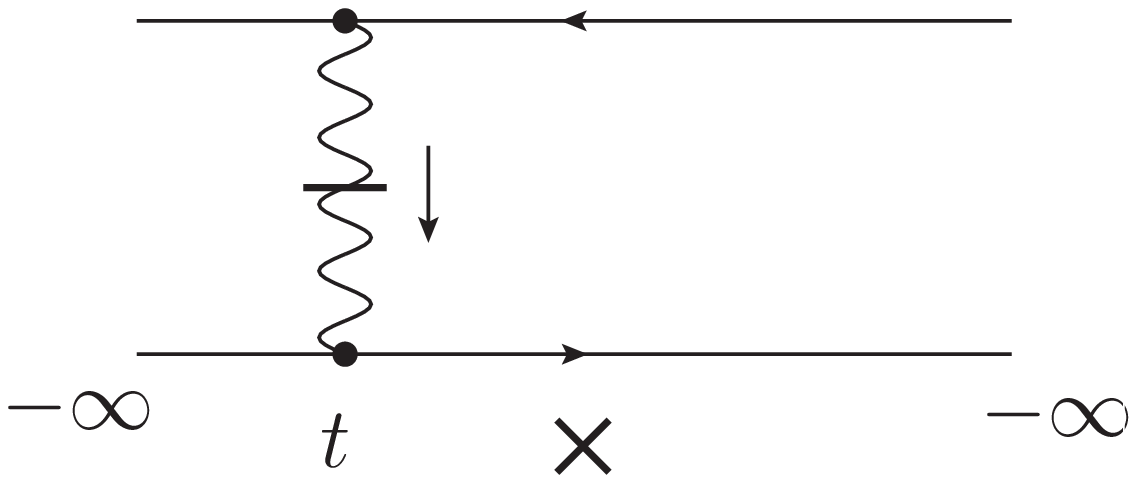}\\
\includegraphics[width=4cm]{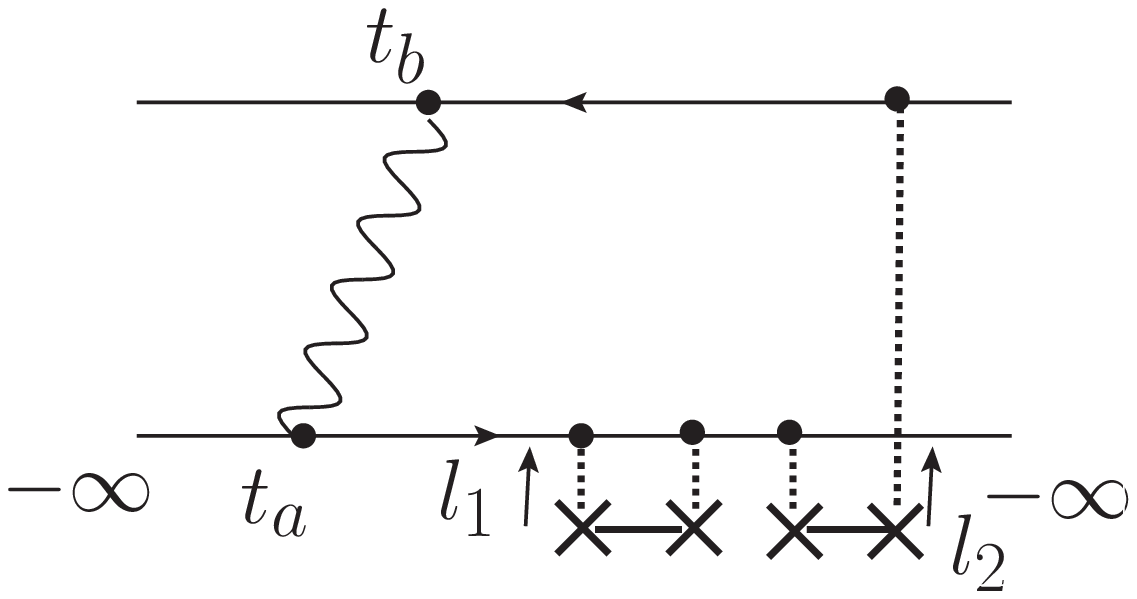}&
\includegraphics[width=4cm]{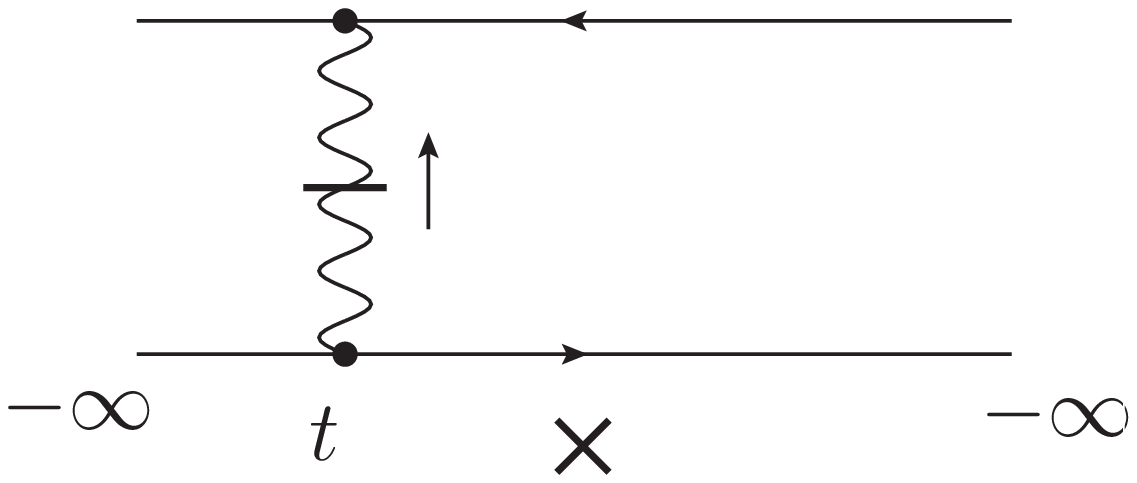}
\end{tabular}
\end{center}
\caption{\label{lo2}
Example of broadening graph (left side of the equality) that corresponds to
a set of 4 graphs on the dipole side (right).
}
\end{figure}

There are however slightly more tricky cases even at leading order. Let us
consider a graph in which the gluon is emitted in the initial state
both in the amplitude and in the complex-conjugate amplitude
(see Fig.~\ref{lo2}).
In this case, one single broadening graph corresponds to 4 graphs on the
dipole side.
This is because the times at which the gluon is emitted in the amplitude
and in the complex-conjugate amplitude respectively are not ordered
on the broadening side. But on the other hand, on the dipole side,
the times at which
the gluon is emitted/absorbed by the quark/antiquark {\em are} ordered,
and different orderings correspond to different lightcone perturbation theory
graphs. Note that
instantaneous-exchange graphs need to be taken into account,
and interestingly enough,
they actually cancel infinities that would plague 
the dipole calculation.
Finally, summing up all the dipole graphs depicted in Fig.~\ref{lo2},
one gets again an exact identification, momentum-by-momentum,
but this time, between one single graph on one side and a set of graphs
on the other side.

The other graphs that one needs to take into account in this leading-order
case are either
symmetric to the ones studied here, or trivial (when, for example,
the gluon couples only to the quark or only to the antiquark on the
dipole side).

All in all, we easily find that the identification between
$p_\perp$-broadening and dipole amplitudes
holds true at leading order.
The identification was first proven, at leading order only,
by Kovchegov et al., see Ref.~\cite{JalilianMarian:2004da} and references therein.


\section{Next-to-leading order}

At next-to-leading order, 
we need to include two radiative gluons. Therefore,
there are many more diagrams that need to be analyzed. 
(The order of magnitude is 100 graphs
on both sides, although we afford to
limit ourselves to the large-$N_c$ limit in order to get rid of all
nonplanar graphs).
Note that the two gluons are assumed to have comparable momenta.
(If the momenta were ordered, the discussion would boil down to 
an iteration of the
leading-order case).

\begin{figure}
\begin{center}
\includegraphics[width=4.5cm]{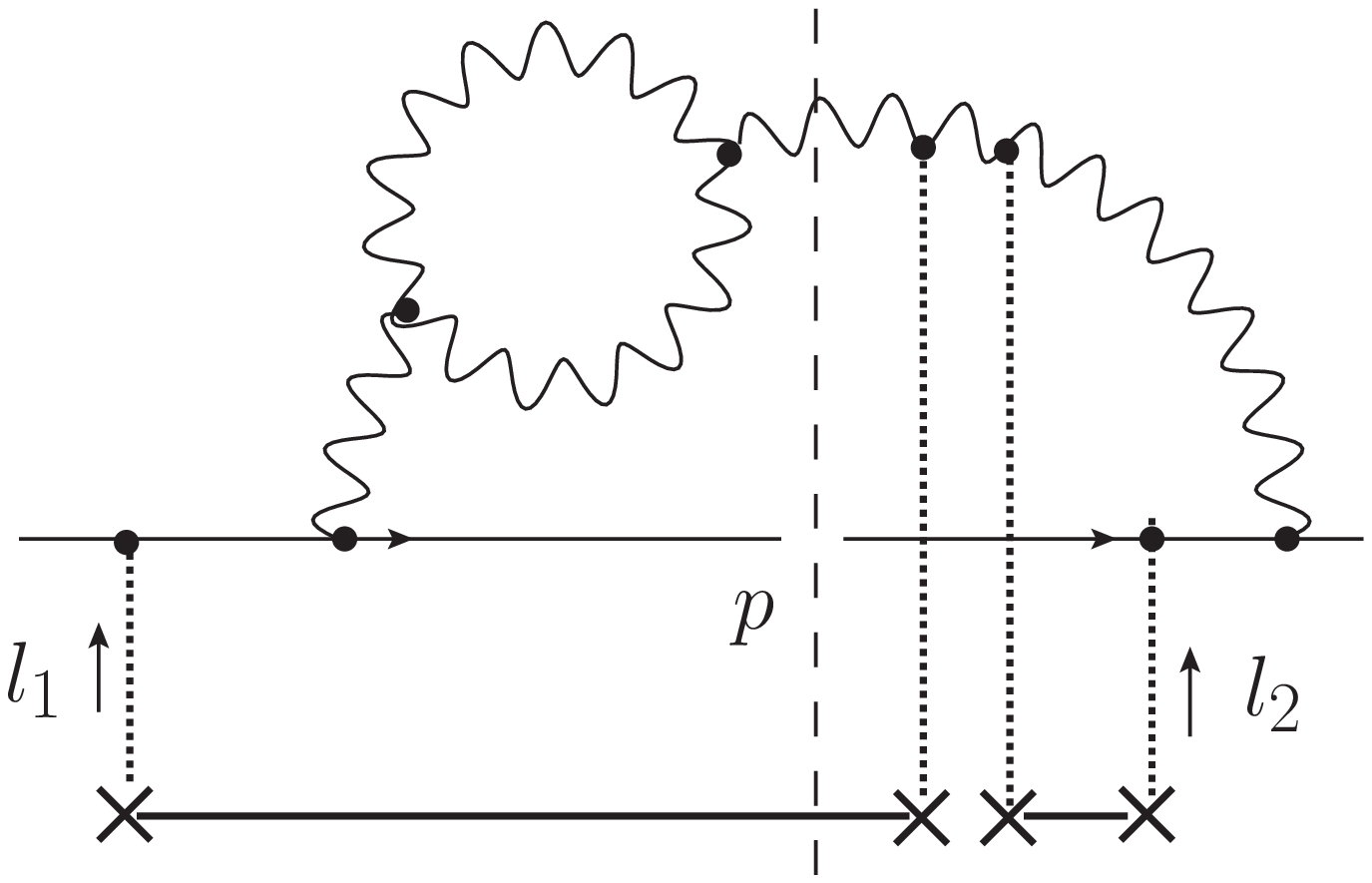}\ \vline\
\includegraphics[width=4.5cm]{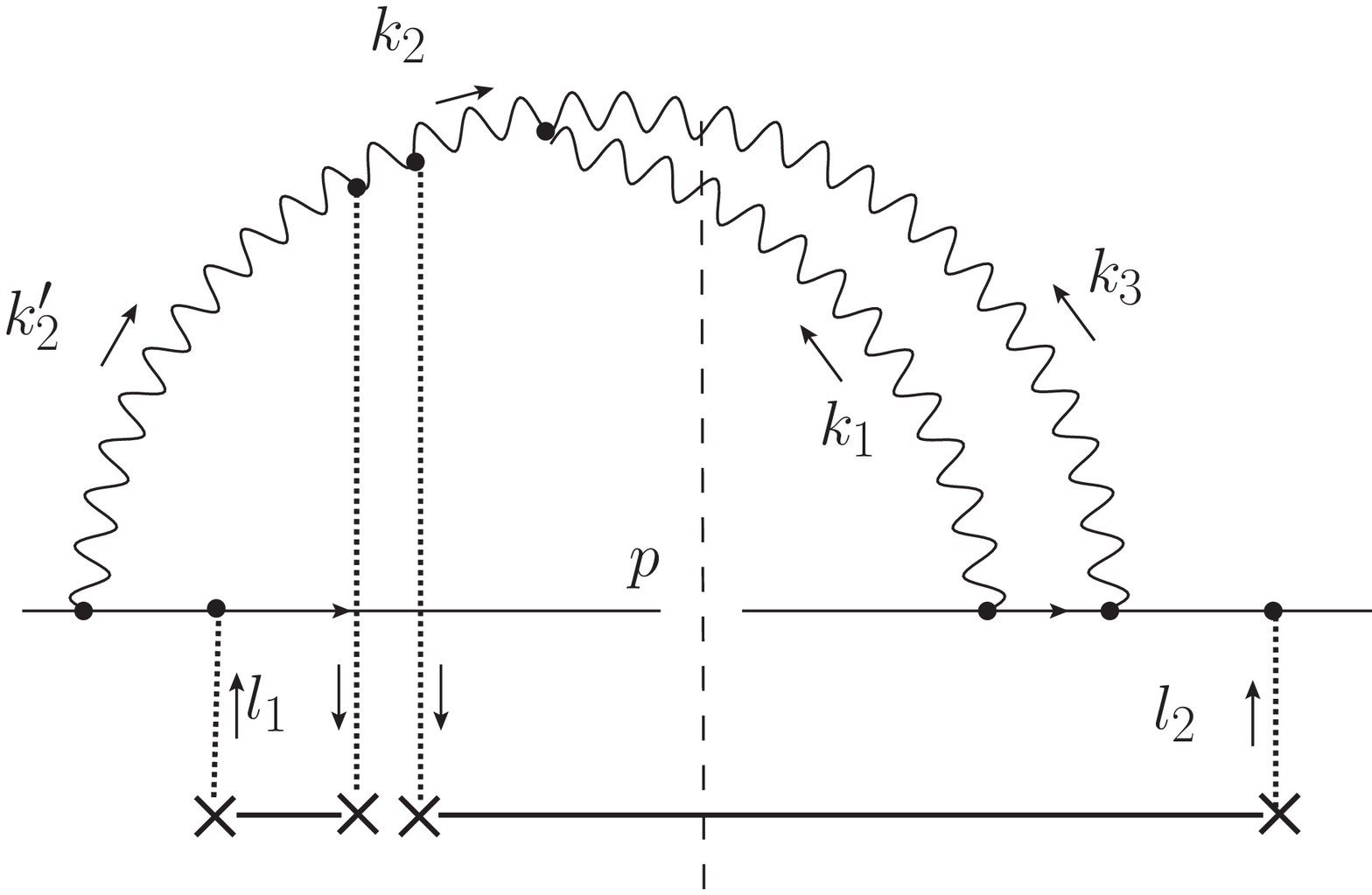}\ \vline\
\includegraphics[width=4.5cm]{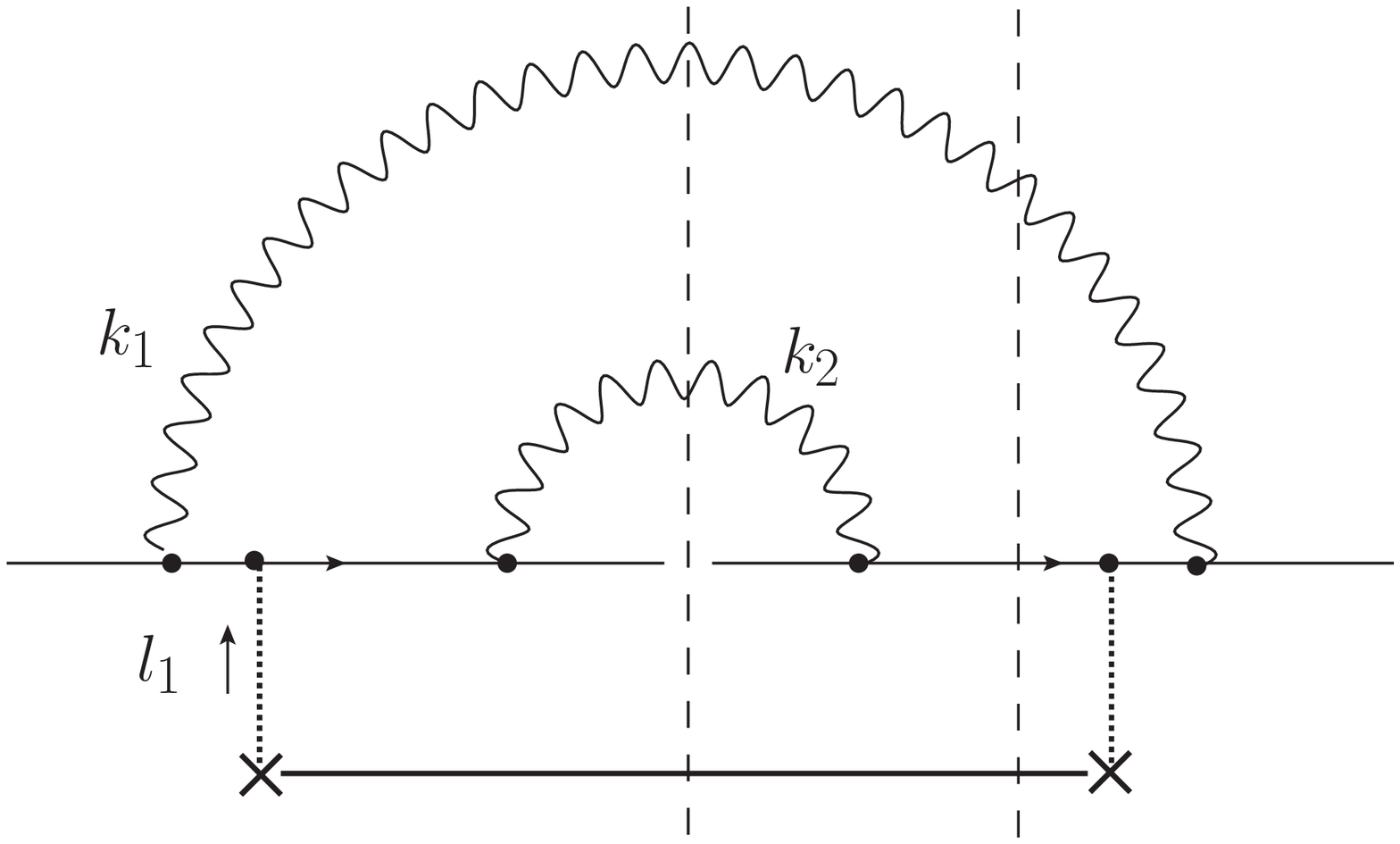}
\end{center}
\caption{\label{NL}One particular representant of each of 
the three main classes of graphs contribution to $p_\perp$-broadening at
next-to-leading order. The radiative gluons carry momenta
of the same order of magnitude.}
\end{figure}

We can classify the graphs in basically 3 classes, according
to the number of quark-gluon vertices (see Fig.~\ref{NL}).
The first class (2 $qg$ vertices)
is actually the easiest one, since it turns out that
we never need to write down the detailed expression for the gluon loop.
Therefore, the equivalence with the similar-looking dipole graphs is
not much more difficult for this class of graphs as for the leading order.
As for the second class (3 $qg$ vertices), 
its analysis requires to write down the full
expression of the exact 3-gluon vertex.
Moreover, for the second and third classes (4 $qg$ vertices), 
the time-ordering problem
already encountered at leading order is not so straightforward to solve.
In order for the ``broadening-dipole identification'' to
become manifest, one actually needs to write the integration over some longitudinal
momentum, go to the complex plane and appropriately
deform the contour, namely, perform an analytical continuation.

Note that we do not need to actually evaluate the momentum integrals
in order to see that the two processes are related.
Such a complete calculation would amount to computing
the BFKL kernel at next-to-leading order in the gluon sector.

\section{Conclusion and outlook}

In the context of proton-nucleus (made of $A$ nucleons) scattering,
we proved that the identity between the rate of production
of a jet of transverse momentum $p_\perp$ and the Fourier transform
of the total cross section for the scattering of a dipole of transverse size
$x_\perp$ conjugate to $p_\perp$ holds true also when two additional 
(unobserved) gluons are included. The gluons are soft compared to the
quark, but no further assumption is required
on the kinematics of these gluons.
The proof holds, strictly speaking, in the large-$N_c$
and large-$A$ limits but the result may eventually be found to hold more
generally. 
By iterating our
calculation to all orders, we actually proved that the two observables
obey the same evolution equation with the rapidity,
namely the Balitsky-Kovchegov (BK) equation at next-to-leading order accuracy.

What we have argued for $p_\perp$-broadening may be generalized 
straightforwardly
to other processes: For example dijet correlations are expected to have
the same evolution as dipole and quadrupole total cross sections.

Our method was a brute-force inspection of all relevant light-cone perturbation theory diagrams. 
Finding a more general and more synthetic
method would be an interesting
and useful challenge. 
Also, we do not know whether the identity would be verified
beyond next-to-leading order accuracy.

\end{document}